\documentclass{article}

\usepackage[margin=1.5in]{geometry}
\usepackage{amsfonts, amsmath, amsthm}
\usepackage[usenames,dvipsnames]{xcolor}
\usepackage[boxruled,commentsnumbered,vlined,linesnumbered]{algorithm2e}
\usepackage{framed}
\usepackage{subcaption}

\usepackage{import}
\usepackage{xparse}
\usepackage{enumitem}
\usepackage{etoolbox}

\definecolor{DarkRed}{rgb}{0.5,0.1,0.1}
\definecolor{DarkBlue}{rgb}{0.1,0.1,0.5}
\usepackage[
  linktocpage=true,
  pagebackref=true,
  colorlinks,
  urlcolor=black,
  linkcolor=DarkRed,
  citecolor=ForestGreen,
  bookmarks,
  bookmarksopen,
  bookmarksnumbered]{hyperref}
\usepackage[noabbrev,nameinlink]{cleveref}

\newtheorem{theorem}{Theorem}[section]
\newtheorem{corollary}{Corollary}[theorem]
\newtheorem{lemma}[theorem]{Lemma}

\newtheorem{observation}[theorem]{Observation}
\newtheorem{definition}{Definition}[section]

\let\eps\varepsilon{}
\DeclareMathOperator{\polylog}{polylog}
\DeclareMathOperator{\supp}{supp}

\AtBeginDocument{%

  \newcommand{\mathsc}[1]{\textnormal{\textsc{#1}}}
  \newcommand{\OPT}{\ensuremath{\mathsc{OPT}}}
  \newcommand{\nonneg}{\mathbb{R}_+}
  \newcommand{\R}{\mathbb{R}}
  
  \newcommand{\Z}{\mathbb{Z}}

  \newcommand{\norm}[1]{\left\lVert#1\right\rVert}
}

\DontPrintSemicolon{}
\SetKwSty{}
\SetStartEndCondition{ }{}{}
\SetKwFor{For}{For}{\string,}{}
\SetKwFor{ForEach}{For each}{}{}
\SetKwIF{If}{ElseIf}{Else}{if}{}{else if}{else}{}
\SetKwProg{Fn}{}{:}{}
\SetFuncSty{textsc}
\SetAlgoHangIndent{0pt}
\SetNlSty{}{}{.}
\SetAlgoNlRelativeSize{0}
\SetKwComment{tcp}{$/\!/$\,}{}

\SetCommentSty{commentfont}
\SetAlgoInsideSkip{smallskip}
\SetKwInOut{Input}{\textbf{Input}}
\SetKwInOut{Output}{\textbf{Output}}
\SetKwProg{proc}{}{:}{}

\crefalias{AlgoLine}{line}%

\makeatletter
\let\cref@old@stepcounter\stepcounter
\def\stepcounter#1{%
  \cref@old@stepcounter{#1}%
  \cref@constructprefix{#1}{\cref@result}%
  \@ifundefined{cref@#1@alias}%
    {\def\@tempa{#1}}%
    {\def\@tempa{\csname cref@#1@alias\endcsname}}%
  \protected@edef\cref@currentlabel{%
    [\@tempa][\arabic{#1}][\cref@result]%
    \csname p@#1\endcsname\csname the#1\endcsname}}
\makeatother
\crefname{algocfline}{line}{lines}
\Crefname{algocfline}{Line}{Lines}

\title{All-Norm Load Balancing in Graph Streams via the Multiplicative Weights Update Method}
\author{Sepehr Assadi \and Aaron Bernstein \and Zachary Langley}

\begin{document}

\begin{titlepage}

\maketitle
\thispagestyle{empty}

\begin{abstract}
  In the weighted load balancing problem, the input is an $n$-vertex bipartite graph between a set of clients and a set of servers, and each client comes with some nonnegative real weight.
  The output is an assignment that maps each client to one of its adjacent servers, and the load of a server is then the sum of the weights of the clients assigned to it.
  The goal is to find an assignment that is well-balanced, typically captured by (approximately) minimizing either the $\ell_\infty$- or $\ell_2$-norm of the server loads.
  Generalizing both of these objectives, the \emph{all-norm load balancing problem} asks for an assignment that approximately minimizes \emph{all} $\ell_p$-norm objectives for $p \ge 1$, including $p = \infty$, simultaneously.

  \smallskip

  Our main result is a deterministic $O(\log{n})$-pass $O(1)$-approximation semi-streaming algorithm for the all-norm load balancing problem.
  Prior to our work, only an $O(\log{n})$-pass $O(\log{n})$-approximation algorithm for the $\ell_\infty$-norm objective was known in the semi-streaming setting.

  \smallskip

  Our algorithm uses a novel application of the multiplicative weights update method to a mixed covering/packing convex program for the all-norm load balancing problem involving an infinite number of constraints.
\end{abstract}

\end{titlepage}

\section{Introduction}

In the \emph{load balancing} problem, the input is a bipartite graph $G=(C, S, E)$ between \emph{clients} $C$ and \emph{servers} $S$ together with a client weight function $w: C \to \mathbb{Z}^+$.
The goal is to find an assignment---a mapping that sends each client to one of its neighboring servers---that is as ``balanced'' as possible.
The \emph{load} of a server under an assignment is the sum of the weights of the clients assigned to it, and the \emph{load vector} is then the vector of server loads, indexed by server.
The notion of ``balance'' can be captured mathematically with a suitable norm of the load vector---the smaller the norm, the more balanced the assignment.
For example, the often considered ``min-max objective''---that of minimizing the maximum server load---is captured by minimizing the $\ell_\infty$-norm.
Though natural, the min-max objective is not always appropriate, as it may neglect to properly balance the clients that are not part of the main bottleneck.
Hence another common objective is to minimize the $\ell_2$-norm.

Generalizing both the $\ell_\infty$- and $\ell_2$-norm objectives, one may ask for an assignment which minimizes the $\ell_p$-norm of the load vector for all $p \ge 1$, including $p = \infty$, simultaneously.
In general, such an assignment may not exist~\cite{AAWY97}, but, perhaps surprisingly, there is always an assignment that is a 2-approximation for every $\ell_p$-norm simultaneously~\cite{AERW04} (see also~\cite{HLLT06,BKPPS17,CS19}).
The problem of computing an assignment that is approximately optimal for all $\ell_p$-norms simultaneously is called the \emph{all-norm load balancing problem}.

We study the all-norm load balancing problem in the semi-streaming model~\cite{FeigenbaumKMSZ04}: the edges of an $n$-vertex graph arrive one-by-one in a stream, and the algorithm is allowed to process these edges using $O(n \polylog{n})$ words of memory and a small number of passes over the stream.
These constraints capture several challenges of processing massive graphs, including I/O-efficiency and efficiently monitoring evolving graphs.
As such, the semi-streaming model has attracted significant attention in recent years.

The load balancing problem has been studied extensively under different terms across various models.
It has a rich history in the scheduling literature as {job scheduling with restricted assignment}~\cite{H73,BCS74,LL04,HLLT06}.
In the distributed computing literature, it has been studied as the backup placement problem~\cite{HKPR18,OBL18,BO20,AssadiBL20} (see also~\cite{SSK06,MT08,KSTZ04,STZ04a,STZ04b} for similar models).
Load balancing has also been studied under the term the semi-matching~\cite{HLLT06,CHSW12,FLN14,KR16}, and from the optimization perspective, it is one of canonical examples of mixed packing-covering optimization problems.

Load balancing can be seen as a cross between maximum matching and coverage problems, which are arguably the two most studied optimization problems in the semi-streaming model (see~\cite{FeigenbaumKMSZ05,McGregor05,AhnG11,GoelKK12,KonradMM12,Kapralov13,CrouchStubbs14,Kapralov21,AssadiLT21,FischerMU21,AssadiJJST22,Bernstein20,AssadiB21,BernsteinDL21} and references therein for matching and~\cite{Assadi17,BEM17,LMSV11,MV17} for coverage---this is by no means a comprehensive list of prior results).
Yet, surprisingly, not much is known about the load balancing problem in this model.
To our knowledge, the only previous work on load balancing in the semi-streaming model, due to Konrad and Rosén~\cite{KR16}, provided a single-pass $O(\sqrt{n})$-approximation algorithm and an $O(\log{n})$-pass $O(\log{n})$-approximation algorithm, both only for the min-max objective and unweighted clients.%
\footnote{
  A very recent work of~\cite{BhattacharyaKS22} also studies mixed packing-covering linear programs in the semi-streaming model.
  While the paper does not explicitly address the load balancing problem, their results seem to imply an $O(1)$-approximation for the (unweighted) min-max objective in $O(\log{n})$ passes~\cite{Sp22}.
  We discuss this connection later in our paper.
}
Our goal in this paper is to address this shortcoming.

\subsection{Our results}
Our main result is the following theorem.

\begin{theorem}[Formalized in~\Cref{thm:main}]\label{thm:main-intro}
  There exists a deterministic semi-streaming algorithm for the all-norm load balancing problem with weighted clients that achieves an $O(1)$-approximation%
  \footnote{
    More precisely, we prove that our algorithm obtains a $19$-approximation.
    We make little attempt to optimize the constant.
    It seems plausible that the approximation factor can be reduced to $8$, but
    not below $4$ without substantial modification.
  }
  in $O(\log{n})$ passes.
\end{theorem}

\Cref{thm:main-intro} vastly generalizes the semi-streaming algorithm of~\cite{KR16} to both weighted clients and to the all-norm guarantee.
At the same time, it improves the approximation ratio from $O(\log{n})$ to $O(1)$.

It is worth pointing out that even for the unweighted min-max objective, obtaining a better-than-$2$-approximation is at least as hard as finding a perfect matching in the input graph.
Indeed, the min-max objective value is one if and only if there is a client-perfect matching in the graph, and otherwise, it is at least two.
Computing a perfect matching requires at least $\Omega(\log{n}/\log\log{n})$ passes in the semi-streaming model~\cite{GuruswamiO16,AssadiR20,ChenKPSY21}, and the current best upper bound is $n^{3/4+o(1)}$ passes~\cite{AssadiJJST22}.

For weighted clients, finding a polynomial-time better-than-2-approximation algorithm for the min-max objective has remained an open problem for 35 years, and it is NP-hard to obtain a better-than-1.5-approximation~\cite{LST87}.
Indeed, any algorithm that rounds a fractional assignment using the algorithm of~\cite{LST87}, as ours does, cannot beat a 2-approximation, even for just the $\ell_\infty$-norm.
Thus, improving the approximation ratio of our algorithm in~\Cref{thm:main-intro} to below $2$ seems to be out of the reach of the current techniques.

\subsection{Our techniques}
We obtain the algorithm of~\Cref{thm:main-intro} by approximately solving a convex program for the all-norm load balancing via the \emph{multiplicative weights update (MWU)} method.
The MWU method has been a particularly successful optimization technique in the streaming model~\cite{AhnG11,BhattacharyaKS22,IMRUVY17} as it provides a way of boosting a ``feasible-on-average'' solution into a feasible and approximately optimal one.
In order to apply the MWU method, one needs to do design a suitable \emph{oracle} and show that the oracle can be implemented via a semi-streaming algorithm efficiently (see~\Cref{sec:mwu}).

The \emph{min-max} objective for load balancing can be formulated as a mixed packing-covering \emph{linear program (LP)}.
The very recent work of~\cite{BhattacharyaKS22} for solving mixed packing-covering LPs in streaming via MWU can then be used to solve this problem.
However, all-norm load balancing is more challenging; there are infinitely many non-linear constraints, and so the previous MWU-based streaming techniques in~\cite{AhnG11,BhattacharyaKS22} do not apply.

At a high level, we implement our MWU-based algorithm by delegating all infinitely many packing constraints of the convex program for all-norm load balancing to the MWU oracle and use MWU itself only to satisfy the (linear) covering constraints.
Roughly speaking, this requires the oracle in each iteration of MWU to find an assignment of a \emph{large fraction} (but crucially not all) of clients to the servers that is balanced with respect to every $\ell_p$-norm simultaneously.
Implementing this oracle is the main technical ingredient of our work.
We postpone the details of this implementation to the streamlined overview of our approach in~\Cref{sec:overview}.

\section{Preliminaries}

We use $\nonneg$ for the nonnegative real numbers.
If $f : \Omega \to R$ for some ground set $\Omega$, we extend $f$ to subsets $A \subseteq \Omega$ by defining $f(A) = \sum_{a \in A} f(a)$.

Throughout the paper, $G = (C, S, E)$ is a bipartite graph on \emph{clients} $C$ and \emph{servers} $S$, and $w : C \to \R_+$ is a weight function on the clients.
We denote the edge $(c, s) \in E$ by $cs$.
For a vertex $v$, we use $\delta(v) = \{ e \in E : e \ni v \}$ to denote the set of edges incident to $v$.

For a function $x : E \to \R$, we define $x^+$ by $x^+(e) = \max \{ x(e), 0 \}$.

\paragraph*{Assignments}
A \emph{partial assignment} $A \subseteq E$ is a set of edges such that every client has degree at most one in $A$.
We say that a client $c$ is \emph{assigned to $s$ (by $A$)} if $cs \in A$.
For $s \in S$, we write $A^{-1}(s)$ to denote the set of clients assigned to $s$ by $A$.
If every client has degree exactly one in $A$, we simply call $A$ an \emph{assignment}.
Every assignment is also a partial assignment. The \emph{load function} $L_A : S \to \nonneg$ of a partial assignment $A$ is given by $L_A(s) = \sum_{c \in A^{-1}(s)} w(c)$, and for $s \in S$ we call $L_A(s)$ the \emph{load of $s$ (under $A$)}.
The $\ell_p$-norm of a partial assignment $A$ is \[
  \left(\sum_{s \in S} L_A(s)^p\right)^{1/p}.
\]

A function $z : E \to \nonneg$ is a \emph{partial fractional assignment} if $z(\delta(c)) \le 1$ for all $c \in C$.
If $z(\delta(c)) = 1$ for all $c \in C$, then we call $z$ a \emph{fractional assignment}.
The \emph{load function} $L_z$ of a partial fractional assignment $z$ is given by $L_z(s) = \sum_{s \in S} \sum_{c \in N(s)} w(c) z(cs)$.
The $\ell_p$-norm of a fractional assignment $A$ is \[
  \left(\sum_{s \in S} L_z(s)^p\right)^{1/p}.
\]

\paragraph*{\texorpdfstring{$b$}{b}-matchings.}
For any function $b : V \to \mathbb{R}_+$, a \emph{fractional $b$-matching} is a function $x : E \to \nonneg$ such that $x(\delta(v)) \le b(v)$ for all $v \in V$.
The value $b(v)$ is is called the \emph{capacity} of $v$.
A vertex $v$ is \emph{saturated} by $x$ if $x(\delta(v)) = b(v)$.
The \emph{size} of the fractional $b$-matching $x$,  is given by $x(E) = \sum_{e \in E} x(e)$.
We say that $x$ is an $\alpha$-approximate fractional $b$-matching if $\alpha x(E) \ge y(E)$ for every fractional $b$-matching $y$.
We say that $x$ \emph{contains} a $b$-matching $y$ if $x(e) \ge y(e)$ for all $e \in E$, which we also denote by $x \ge y$.

As we work with bipartite graphs on clients and servers, it will often be convenient to specify the vertex capacities of the clients and servers separately.
To that end, for $\kappa : C \to \nonneg$ and $\tau : S \to \nonneg$, a \emph{fractional $(\kappa, \tau)$-matching} is a fractional $b$-matching where $b(v) = \begin{cases}\kappa(v) & v \in C\\\tau(v) & v \in S\end{cases}.$

\paragraph*{All-Norm Load Balancing}

In the \emph{all-norm load balancing problem}, the input is a bipartite graph $G = (C, S, E)$ together with a weight function $w : C \to \Z_+$.
We call elements of $C$ \emph{clients} and elements of $S$ \emph{servers}.
The goal is to find an assignment $A \subseteq E$ that is approximately optimal with respect to every $\ell_p$-norm objective simultaneously.
Indeed, we must settle for an approximation as it is not always possible to minimize all $\ell_p$-norms simultaneously.
But there always exists an assignment whose $\ell_p$-norm is at most twice the optimal $\ell_p$-norm for all $p \ge 1$~\cite{AERW04}.

An assignment $A$ is said to be a \emph{$\beta$-approximation all-norm assignment}, or more simply a \emph{$\beta$-all-norm assignment}, if $\lVert L_A \rVert_p \le \beta \cdot \OPT_p$ for all $p \ge 1$, where $\OPT_p$ is the $\ell_p$-norm of the $\ell_p$-norm minimizing assignment.

\paragraph*{The Multiplicative Weights Update Method}\label{sec:mwu}

The multiplicative weights update (MWU) method is a generic and powerful meta-algorithm that, among other things, can be used to find (approximately) feasible points in a convex set partially defined by halfspaces.

Suppose we have a set of $n$ linear inequalities in $m$ variables $\mathbf{z} \in \R^m$ given by $a_i \cdot z \ge b_i$, where $a_i \in \R^m$ and $b_i \in \R$ for $i \in \{1, \dots, n\}$.
Suppose further there is a convex set $\mathcal{P} \subseteq \R^m$, and we want to find a $z \in \mathcal{P}$ such that $z$ is feasible for the linear inequalities above.
We call the linear inequalities $a_i \cdot z \ge b_i$ the ``hard'' constraints and the constraints that form the convex set $\mathcal{P}$ the ``easy'' constraints.
The MWU method introduces a value $\lambda_i$ for each constraint $i \in [m]$ and reduces the task to the following: find a $z \in \mathcal{P}$ such that $\sum_{i = 1}^m \lambda_i a_i \cdot z \ge \sum_{i = 1}^m \lambda_i b_i$.
Algorithms that solve this simpler task in service to the MWU method are called \emph{oracles}.

Not all oracles are created equal; although the MWU method can use any oracle to obtain a solution to the original problem, oracles that produce solutions that are ``closer'' to being feasible to the actual constraints require fewer invocations.

\begin{definition}
  A \emph{$d$-bounded oracle} is an algorithm which given a vector $\lambda \in \nonneg^m$ computes an $z \in \mathcal{P}$ such that \[
    \sum_{i=1}^m \lambda_i a_i \cdot z \ge \sum_{i=1}^m \lambda_i b_i
  \]
  or correctly reports that no such $z$ exists.
  Moreover, it holds that $|a_i z - b_i| \le d$ for all $i \in [m]$.
\end{definition}

\begin{algorithm}
  \caption{The multiplicative weights update (MWU) method.}\label{alg:mwu}
  \SetKwInput{Given}{Given}

  \Given{A $d$-bounded oracle for the feasibility problem and an error parameter $\eps \in (0, 1)$.}

  Set $\eta \gets \eps / (4d)$.
  Set $T \gets 8 d^2 \log{n} / \eps^2$.
  Set $\lambda_i^{(1)} \gets 1$ for each constraint $i \in [n]$.\;
  \For{$t = 1$ \KwTo $T$}{
    Run the given $d$-bounded oracle with $\lambda^{(t)}$ to obtain solution $z^{(t)}$. If the oracle returns \emph{infeasible}, then return \emph{infeasible}.\;
    Set $m_i^{(t)} \gets \frac{1}{d}(b_i - A_i z^{(t)})$ for each $i \in [n]$.\;
    Set $\lambda_i^{(t + 1)} \gets \lambda_i^{(t)} (1 - \eta m_i^{(t)})$ for each $i \in [n]$.\;
  }
  \Return{$\frac{1}{T} \sum_{t=1}^T z^{(t)}$}\;
\end{algorithm}

The following theorem follows from \cite{AHK12} and the fact that \Cref{alg:mwu} can be implemented in the streaming setting if given an oracle for such a setting.

\begin{theorem}\label{thm:mwu}
  Let $\eps > 0$ be an error parameter.
  Suppose we have a stream defining the constraints $a_i z \ge b_i$.
  Further suppose that there exists a $d$-bounded oracle for the feasibility problem that makes $p$ passes over the stream and uses $s$ space.
  Then \Cref{alg:mwu} equipped with this oracle either returns an $x$ such that $a_i \cdot x \ge b_i - \eps$ or correctly decides that the system is infeasible.
  The algorithm makes $O(pd^2 \log{m}/\eps^2)$ passes over the stream and uses $O(n + sd^2\log{m}/\eps^2)$ space.
\end{theorem}

\newcommand{\allCj}[1][j]{\ensuremath{(C^{(#1)})_{j=0}^{\jmax}}}
\newcommand{\Cj}[1][j]{\ensuremath{C^{(#1)}}}
\newcommand{\Ej}[1][j]{\ensuremath{E^{(#1)}}}
\newcommand{\Gj}[1][j]{\ensuremath{G^{(#1)}}}
\newcommand{\Cmax}{C^{(\jmax)}}

\newcommand{\lpcostz}{%
  \left(\sum_{s \in S} \bigg(\sum_{c \in N(s)} w(c) z(cs)\bigg)^p\right)^{1/p}%
}

\newcommand{\imax}{\ell}%{{i_{\sf max}}}
\newcommand{\jmax}{k}%{{j_{\sf max}}}

\section{High-Level Overview}\label{sec:overview}

Our goal is to compute an $O(1)$-all-norm assignment in the semi-streaming setting using $O(\log{n})$ passes over the stream.
Our approach can be viewed as the following series of reductions:
\begin{enumerate}
  \item Computing an $O(1)$-all-norm assignment reduces to computing an $O(1)$-all-norm \emph{fractional} assignment.
  \item Computing an $O(1)$-all-norm fractional assignment reduces, via the MWU method, to solving a problem we call the \emph{$O(1)$-all-norm oracle problem}.
  \item The $O(1)$-all-norm oracle problem in turn reduces to computing a \emph{matching hierarchy with size factor $O(1)$}.
  \item Finally, computing a matching hierarchy with size factor $O(1)$ reduces to computing $O(\log^2{n})$ maximal $b$-matchings.
\end{enumerate}
The algorithm providing the reduction for the first step is well-known and is due to Lenstra, Shmoys and Tardos~\cite{LST87}.
The second reduction is provided by the multiplicative weight framework applied to the infinite-constraint convex program that captures the $O(1)$-all-norm fractional assignment problem.
The third and fourth steps are our main algorithmic contributions.
Most of the complexity is in the third step, and this section is devoted to providing intuition for that step.

\paragraph*{The $O(1)$-all-norm oracle problem.}
In the \emph{$\beta$-all-norm oracle problem}, the input is a graph $G = (C, S, E)$, weights $w : C \to \nonneg$, and \emph{values} $\lambda : C \to \nonneg$.
The goal is to compute a function $z : E \to \nonneg$ such that \[
  \sum_{c \in C} \lambda(c) z(\delta(c)) \ge \sum_{c \in C} \lambda(c)
\]
and \[
  \lpcostz \le \beta \cdot \OPT_p
\]
for all $p \ge 1$, where $\OPT_p$ is the $\ell_p$-norm of the $\ell_p$-norm minimizing assignment.
We say that an algorithm solves the \emph{$O(1)$-all-norm oracle problem} if it solves the $\beta$-all-norm oracle for some constant $\beta$.

Let us assume $w(c)=1$ for all $c \in C$.
This assumption is not without loss of generality, but since it already captures many of the key algorithmic ideas in our approach, we make this assumption to simplify the following discussion.

It suffices to find a partial assignment $A \subseteq E$ such that
\begin{equation}\label{eq:overview-ccc}
  \sum_{cs \in A} \lambda(c) \ge \frac{1}{\alpha} \sum_{c \in C} \lambda(c).
\end{equation}
for some $\alpha \ge 1$ and \begin{equation}\label{eq:overview-lpc}
  \left(\sum_{s \in S} |L^{-1}(a)|^p\right)^{1/p} \le O(1) \cdot \OPT_p.
\end{equation}
Indeed, then scaling the indicator of $A$ by $\alpha$ yields a solution.

\paragraph*{A charging scheme for partial assignments.}
Let $A \subseteq E$ be a partial assignment and let $\prec$ be an arbitrary total ordering of $C$.
We now describe a useful charging scheme that allocates the quantity $\lVert L_A \rVert_p^p$ to the clients according to $\prec$.
More formally, we define a function $\phi_p : C \to \nonneg$ for each $p \ge 1$ with the property that
\begin{equation}\label{eq:charge}
  \sum_{c \in C} \phi_p(c) = \lVert L_A \rVert_p^p.
\end{equation}
The definition is simple: If $A$ does not assign $c$, set $\phi_p(c) = 0$.
Otherwise, if $c$ is the $k$th smallest element (with respect to $\prec$) among the clients assigned to $A(c)$, then set $\phi_p(c) = k^p - (k - 1)^p$.
It is easy to see that $\sum_{c \in A^{-1}(s)} \phi_p(c) = L_A(s)^p$ as the sum telescopes.
Hence $\phi_p$ satisfies \eqref{eq:charge} as desired.

This charging scheme will be the basis for an upper bound on the cost of the solution our algorithm computes and also for a lower bound on the optimal cost.

\paragraph*{Matching hierarchies and $\ell_p$-norms.}
To solve the $O(1)$-all-norm oracle problem---that is, to find an $A \subseteq E$ that satisfies \eqref{eq:overview-ccc} and \eqref{eq:overview-lpc}---we first compute a \emph{matching hierarchy}.
Let $\ell = \lceil \log_2 n \rceil$.
A matching hierarchy is a sequence of matchings $M_0, \dots, M_\ell$ such that $M_i$ is a $(1, 2^i)$-matching and $M_0 \subseteq \cdots \subseteq M_\ell$.
By definition, if a client $c$ is matched to a server $s$ in some $M_i$, then it is matched to $s$ in $M_{i'}$ for all $i' \ge i$.
Let the \emph{rank} of a client $c$ be the smallest $i$ such that $c$ is matched in $M_i$ or $\infty$ if no such $i$ exists.

A matching hierarchy provides two main benefits.
First, it gives us a foundation from which to build a partial assignment.
Notice that each $M_i$ is a partial assignment of maximum load $2^i$, and because the $M_i$'s are nested, every $M_i$ that assigns a client $c$ agrees on which server to assign $c$ to.
In the end, our final solution $A$ will be a subset of $M_\ell$; that is, we will assign some clients according to $M_\ell$ and leave others unassigned.

The second benefit of a matching hierarchy is what differentiates it from a simple assignment.
A matching hierarchy naturally defines a useful ordering of the clients, and any ordering that lists the clients in non-decreasing order of rank will do.
To see why such an ordering is useful, consider an $A \subseteq M_\ell$. If $c$ is assigned by $A$ and $c$ has rank $i$ (in the matching hierarchy) then $c$ must appear in the first $2^i$ clients assigned to $s$ in $A$.
It follows that $\phi(c) \le 2^{ip} - (2^i - 1)^p.$
Hence the $\ell_p$-norm of $A \subseteq M_\ell$ is upper bounded by a function of the ranks of the clients as defined by the matching hierarchy.

We can also obtain a lower bound for $\OPT^p_p$ in terms of a matching hierarchy.
Let $A^*$ be an $\ell_p$-norm minimizing assignment and consider an arbitrary ordering of the clients.
For $i \in \{0, \dots, \ell \}$, define $M^*_i$ to be the $(1, 2^i)$-matching that assigns to each server its first $2^i$ clients.
Now consider the same charging scheme as before, this time applied to $M^*_\ell$.
A client $c$ of rank $0$, by the definition of $M^*_0$, is the first element of its server's preimage under $A$, and hence $\phi(c) = 1$.
A client $c$ of rank $i$ is preceded by at least $2^{i - 1}$ clients in its server's preimage under $A$, and hence $\phi(c) \ge (2^{i-1} + 1)^p - 2^{(i-1)p}$.

The main takeaway is the following.
The matching hierarchy $A$ that we compute assigns a rank to each client, and we can also decompose an optimal solution into a (different) matching hierarchy that defines its own rank on each client.
To prove that $A$ is near-optimal, we compare the ranks of the clients in $A$ to those in the optimal solution.

\paragraph*{Size factor.}
It is not enough to merely have many clients of low rank; we also need to make sure we have assigned clients with enough $\lambda$-value, or else we may not satisfy \eqref{eq:overview-ccc}.
One approach would be to ensure that the $\lambda-$value matched by each $M_i$ is a constant fraction of the $\lambda$-value matched by any other $(1, 2^i)$-matching.
We will want something a bit stronger.

Let $\Cj = \{ c : \lambda(c) \ge \max_{c \in C} \lambda(c) \cdot 2^{-j} \},$ let $\Ej = \{ cs \in E : c \in \Cj \}$, and let $\Gj = (C, S, \Ej)$.
We say that a matching hierarchy $M_0, \dots, M_\ell$ has \emph{size factor $\alpha$} if $M_i \cap \Ej$ is an $\alpha$-approximate maximum cardinality $(1, 2^i)$-matching of $\Gj$ for all $i, j \in \{0, \dots, \log{n}\}$.

Notice that having size factor $\alpha$ implies, in particular, that the value matched by each $M_i$ is at least a $1/\alpha$ fraction of the value matched by any other $(1, 2^i)$-matching.
That is, $M_i$ is an $\alpha$-approximate maximum ``weight'' matching where the ``weight'' of an edge $cs$ is given by $\lambda(c)$ (the quotation marks are to emphasize that this ``weight'' is different from the weight $w(c)$ of a client $c$).

To see why such a matching hierarchy is useful consider the following.
We want to show that compared to an optimal matching hierarchy, our solution has many low-rank clients of high cost.
Pick any rank-threshold $i$ and note that that the maximum $(1,2^i)$-matching in $\Gj$ is precisely the number of clients that have rank $\leq i$ in $A^*$ and value $\geq \max_{c \in C} \lambda(c) 2^{-j}$.
Thus, a matching hierarchy of size factor $\alpha$ has at least $1/\alpha$ as many clients as $A^*$ of this category.
In other words, for any rank/cost tradeoff, the matching hierarchy is competitive with $M_0^*, \dots, M_\ell^*$.

\paragraph*{Approaches that fail.}
Once we compute a matching hierarchy $M_0, \dots, M_\ell$, we then need to compute a partial assignment satisfying \eqref{eq:overview-ccc} and \eqref{eq:overview-lpc}.
One naive solution is to let the partial assignment be $M_\ell$ itself.

To see why this fails, consider the following example.
Let $n$ be a power of two, let $C = \{c_1, \dots, c_n\}$ and let $S = \{s_1, \dots, s_n\}$.
The graph $G$ is the complete graph on $C$ and $S$, and $\lambda(c) = 1$ for all $c \in C$.
The nested matching hierarchy is the following.
The matching $M_0$ matches $c_k$ to $s_k$ for $k \in \{1, \dots, n/2\}$.
For $i \ge 1$, each $M_i$ builds on $M_{i-1}$ by assigning $2^{i-1}$ previously unassigned clients to $s_1$.
As every $M_i$ matches the first $n/2$ clients, the matching hierarchy clearly has size factor 2.
But $s_1$ has load $n/2$ in $M_\ell$, while the $\ell_\infty$-norm of the optimal assignment is one ($G$ has a perfect matching).

Thus, instead of picking all of $M_\ell$, the idea is to let $A$ assign some strict subset of the clients according to $M_\ell$ and leave the others unassigned.
In light of the above example, one other tempting option is to simply pick clients in increasing order of rank until we have collected enough $\lambda$-value.
However, this approach fails when there are many low-rank clients with low value.

%The example is similar to the previous one, except we augment the graph with a new server adjacent to $n$ new clients, all with value $n$.
%The matching hierarchy is the same as before, except each $M_i$ also includes $2^i$ of the new edges.
%The proposed algorithm will need to include all edges of $M_{\ell-2}$ before it accumulates enough value.
%In particular, there will be a server of load $n/4$ (from the star) and one of load $\log_2{n} - 2$ (from the complete graph).
%The $\ell_2$ norm is thus at least $\sqrt{n^2 + \log_2{n}^2}$.
%But the optimal load vector is $(n, 1, \dots, 1)$ and has $\ell_2$-norm $\sqrt{2n}$.

The problem with the approach in the previous paragraph is that it does not account for the value of the clients.
An equally myopic approach would be to ignore rank and assign clients in decreasing order of value until \eqref{eq:overview-ccc} is satisfied.
Our first example already shows why this does not work.

Thus when picking the subset clients we need to maintain a balance between the cost of the clients and their value.
We already have a good proxy for the cost of a client with respect to an $\ell_p$-norm: if we set $\hat{\phi}_p(c) = 2^{ip} - (2^i - 1)^p$ where $i$ is the rank of $c$, then $\hat{\phi}_p(c) \le \phi_p(c)$.
We can therefore assign to each client then a cost and a value, and the problem then resembles a fractional knapsack problem.
It is natural therefore to try a fractional-knapsack-like algorithm: in decreasing order of $\hat{\phi}_p(c) / \lambda(c)$, assign the clients greedily until \eqref{eq:overview-ccc} is satisfied.
Indeed, if one is only interested in minimizing a \emph{fixed} $\ell_p$-norm, this approach can be made to work.
The problem for all-norm load balancing is that $\hat{\phi}_p$ is a function of $p$, and the ordering it induces depends on the choice of $p$.

\paragraph*{Our algorithm.}
Our algorithm prioritizes value in the following way.
First it only considers clients in $\Cj[0]$.
The algorithm begins collecting clients from $\Cj[0]$ in increasing order of rank.
The next step is crucial.
Once the algorithm has collected \emph{exactly} $\frac{1}{\alpha} |\Cj[0]|$ clients,%
\footnote{
  We assume for simplicity that $\alpha$ divides $|\Cj|$. Our actual algorithm
  computes a fractional partial assignment, so the issue of divisibility
  vanishes.
}
it ends the $\Cj[0]$ iteration.
The algorithm then expands the set of clients under its consideration to $\Cj[1]$ and resumes adding clients (that it has not yet taken) in increasing order of rank.
Again, once it has collected \emph{exactly} $\frac{1}{\alpha} |\Cj[1]|$ clients, it proceeds to $\Cj[2]$.
And so on.

It is not hard to see that the algorithm will collect at least a $1/\alpha$ fraction of $\lambda(C)$. Indeed, by first collecting $|\Cj[0]|/\alpha$ clients from $\Cj[0]$, then collecting $\Cj[1]/\alpha$ clients from $\Cj[1]$, and so on, the algorithm will end with a $1/\alpha$ fraction of the total value.

Matching hierarchies provide us with the key tool for proving that an assignment is near-optimal for \emph{all} $\ell_p$-norms.
For every $i,j$, let $N_{i,j}$ denote the number of clients in our hierarchy of rank $\le i$ and value $\ge \max_{c \in C} \lambda(c) 2^{-j}$, and let $N^*_{i,j}$ be the corresponding number for an optimal matching hierarchy.
If we could guarantee that for all $i,j$ we have $N_{i,j} \sim N^*_{i,j}$ then it is not hard to check that our solution would be approximately all-norm optimal (consider the charging scheme described above).
We do not quite achieve $N_{ij} \sim N^*_{ij}$, but we do achieve a slightly weaker guarantee which is still sufficient for approximate all-norm optimality.

\paragraph*{Weighted clients.}
There is well-known albeit inefficient reduction from weighted fractional load balancing to unweighted load balancing: Replace every client $c$ with $w(c)$ replicas of unit weight and give each client value $\lambda(c)/w(c)$.
Indeed, any assignment in this client-replicated graph can be mapped back to a fractional assignment in the original graph with the same load vector.
Since every assignment in the weighted graph also corresponds to an assignment in the client-replicated graph with the same $\ell_p$-norm, it follows that a $\beta$-approximate all-norm assignment in the client-replicated graph is a $\beta$-approximate all-norm fractional assignment in the original graph.

Rather than explicitly running our streaming algorithm through this reduction, which, at least if applied naively, does not work in the semi-streaming setting, we instead maintain a fractional assignment directly.
All intuition, however, comes from the idea of client-replication.
For example, in our unweighted algorithm, we consider clients in decreasing order of value.
In our weighted algorithm, emulating the unweighted algorithm on the client-replicated graph, we consider clients in decreasing order of $\lambda(c)/w(c)$.

\newcommand{\Emax}{E^{(\jmax)}}

\newcommand{\Gmax}{G^{(\jmax)}}

\newcommand{\bx}{\ensuremath{\mathbf{x}}}
\newcommand{\by}{\ensuremath{\mathbf{y}}}

\newcommand{\tx}{\ensuremath{\tilde{x}}}
\newcommand{\xj}[1][j]{\ensuremath{x^{(#1)}}}
\NewDocumentCommand{\xji}{ O{j} O{i} }{\ensuremath{x^{(#1,#2)}}}
\newcommand{\txj}[1][j]{\ensuremath{\tx^{(#1)}}}
\newcommand{\xsj}[1][j]{\ensuremath{{x^*}^{(#1)}}}
\newcommand{\xmax}{x^{(\jmax)}}
\newcommand{\txmax}{\tx^{(\jmax)}}

\section{All-Norm Load Balancing in the Streaming Model}\label{sec:streaming}

\subsection{The All-Norm Oracle Problem}

In this subsection, we how show the multiplicative weights framework together with a classic rounding algorithm reduce the task of $O(1)$-all-norm load balancing to that of solving a simpler problem, which we call the $O(1)$-all-norm oracle problem.
Once we establish the reduction, the reader can safely forget about the multiplicative weights update method and focus exclusively on solving the $O(1)$-all-norm oracle problem in the semi-streaming setting.

\paragraph*{Rounding.}
To compute a sparse $O(1)$-all-norm assignment, it is sufficient to compute a sparse $O(1)$-all-norm \emph{fractional} assignment.
Indeed, the classic algorithm of Lenstra, Shmoys, and Tardos~\cite{LST87} does the trick.
\begin{lemma}[\cite{AERW04}]\label{lem:rounding}
  The rounding algorithm of~\cite{LST87} takes as input a fractional assignment $z : E \to \nonneg$ and produces an (integral) assignment $A \subseteq E$ such that $\norm{L_A}_p \le \norm{L_z}_p + \lVert w \rVert_p.$
  Furthermore, the algorithm uses $O(\supp(z))$ space.
\end{lemma}
Hence if we can compute an $O(1)$-all-norm fractional assignment in the semi-streaming setting with support $O(n \polylog{n})$, then we can simply round it as a post-processing step using \Cref{lem:rounding}.

\paragraph*{The $O(1)$-all-norm oracle problem.}
To compute an $O(1)$-all-norm fractional assignment in the semi-streaming setting, we employ the multiplicative weights update framework.
Recall that there are two types of constraints that a $\beta$-approximate all-norm fractional assignment $z : E \to \nonneg$ must satisfy.
First there are the \emph{client covering constraints}, which are the linear constraints \(
  \sum_{s \in N(c)} z(cs) \ge 1
\)
for each $c \in C$.
There are also the \emph{$\beta$-approximate $\ell_p$-norm constraints}, which are the convex constraints \[
  \lpcostz \le \beta \cdot \OPT_p,
\]
for all $p \ge 1$.

To apply the MWU method, we treat the $\beta$-approximate $\ell_p$-norm constraints as defining the convex set $\mathcal{P}$.
Although there are infinitely many such constraints, the set of feasible points forms a convex set, and so the uncountable nature of the constraints poses no problem for the framework.

\begin{figure}[ht]
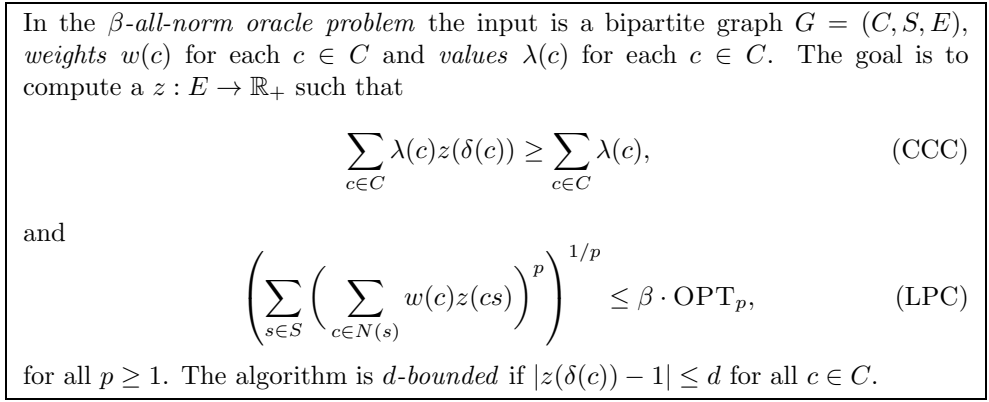

	\centering
	\fbox{
		\begin{minipage}{.9 \textwidth}
			In the \emph{$\beta$-all-norm oracle problem} the input is a bipartite graph $G = (C, S, E)$, \emph{weights} $w(c)$ for each $c \in C$ and \emph{values} $\lambda(c)$ for each $c \in C$.
			The goal is to compute a $z : E \to \nonneg$ such that
			\begin{equation}\label{eq:ccc}
				\sum_{c \in C} \lambda(c) z(\delta(c)) \ge \sum_{c \in C} \lambda(c),\tag{CCC}
			\end{equation}
			and
			\begin{equation}\label{eq:lpc}
				\lpcostz \le \beta \cdot \OPT_p,\tag{LPC}
			\end{equation}
			for all $p \ge 1$.
			The algorithm is \emph{$d$-bounded} if $|z(\delta(c)) - 1| \le d$ for all $c \in C$.
		\end{minipage}
	}
	\caption{The $\beta$-all-norm oracle problem.}
	\label{fig:oracle}
\end{figure}

In this formulation, the MWU method introduces a value $\lambda(c)$ for each client $c$.
The task is then to design an oracle that computes a $z : E \to \nonneg$ such that the $\beta$-approximate $\ell_p$-norm constraints are satisfied for all $p \ge 1$ and
\begin{equation}\label{eq:combined-constraint}
    \sum_{c \in C} \lambda(c) z(\delta(c)) \ge \sum_{c \in C} \lambda(c).
\end{equation}
We call constraint \eqref{eq:combined-constraint} the \emph{combined client covering constraint}.
The problem is formally stated in \Cref{fig:oracle}.

We can now prove the main lemma of this section.

\begin{lemma}\label{lem:oracle-reduction}
    If there is a $d$-bounded $p$-pass $S(n, m)$-space streaming algorithm for the $\beta$-all-norm oracle problem, then there is an $O(p d^2 \beta^2 \log{n})$-pass $O(S(n, m) \cdot \beta^2 d^2 \log{n} + n)$-space streaming algorithm to compute a $(\beta + 2)$-all-norm assignment.
\end{lemma}
\begin{proof}
  Run \Cref{alg:mwu} equipped with the provided oracle and error parameter $\eps = 1/\beta$.
  By \Cref{thm:mwu}, the result $\bar{z}$ satisfies $\bar{z}(\delta(c)) \ge 1 - \eps/2$ for each client $c$ and $\norm{L_{\bar{z}}}_p \le \beta \cdot \OPT_p$ for all $p \ge 1$.
  By scaling $\bar{z}$ by $1 + \eps$ and running the rounding algorithm of \Cref{lem:rounding} on the result, we obtain an integral assignment $\hat{z}$ such that \[
	  \norm{L_{\hat{z}}}_p \le (1 + \eps) \norm{L_{\bar{z}}}_p + \norm{w}_p
		  \le (1 + \eps) \beta \cdot \OPT_p + \OPT_p \le (\beta + 2) \cdot \OPT_p
  \]
  for all $p \ge 1$, as desired.

  Each invocation of the oracle requires $p$ passes over the stream and $S(n, m)$.
  Because the width of the oracle is $d$, the MWU algorithm makes $O(d^2 \log{n}/\eps^2)$ calls to the oracle.
  Hence $O(d^2 p \log{n} / \eps^2)$ passes are used.

  Finally, the space requirements by \Cref{thm:mwu} are $O(S(n,m) \cdot d^2 \log{n}/\eps^2 + n)$, and running the rounding algorithm cannot increase this by more than a factor of two.
\end{proof}

\subsection{Nested Matching Hierarchies}

Recall that in the all-norm oracle problem (\Cref{fig:oracle}), each client $c$ is associated with two quantities, a value $\lambda(c)$ and a weight $w(c)$.
Assigning clients with large value makes it easier to satisfy the combined client covering constraint~\eqref{eq:ccc}, while assigning clients with large weight makes it harder to satisfy the $\ell_p$-norm constraints~\eqref{eq:lpc}.
The \emph{value-per-unit-weight} of a client $c$, denoted by $\rho(c)$, is $\lambda(c)/w(c)$.
Let $\eps > 0$; we will choose $\eps$ later.

\begin{definition}\label{def:vpw}
  The \emph{value-per-unit-weight of a client $c$}, denoted $\rho(c)$ is $\lambda(c) / w(c)$.
  For $j \in \Z$ define $\Cj = \{ c \in C : \rho(c) \ge \max_{c \in C} \rho(c) \cdot e^{-\eps j}\}.$
  Also define $\Ej = \{ cs \in E : c \in \Cj \}$ and $\Gj = (C, S, \Ej)$.
\end{definition}

We use that $\Ej[-1] = \emptyset$ to simplify some of our proofs.
Note that $\Cj[0]$ is the set of clients with \emph{largest} value-per-unit-weight, and also that $\Cj \subseteq \Cj[j+1]$ for all integers $j$.
The next simple lemma says that we can ignore clients outside $\Cj$ for some $j = O(\log(\max_{c \in C} \rho(c) w(C))/\eps)$.

\begin{lemma}\label{lem:lambda-cmax}
  Let $R = \max_{c \in C} \rho(c)$.
  For $j \ge \ln(R \cdot w(C)/\eps) / \eps$, it holds that $\lambda(\Cj) \ge \lambda(C) - \eps.$
\end{lemma}
\begin{proof}
  $\lambda(C) - \lambda(\Cj) = \lambda(C \setminus \Cj) = \sum_{c \in C \setminus \Cj} \rho(c) w(c)
                             \le Re^{-\eps j} w(C \setminus \Cj)
                             \le \eps.$
\end{proof}

A fractional $(\kappa, 2^i$)-nested-matching-hierarchy is a sequence of $b$-matchings $x_0, x_1, \dots, x_\imax$ where each $x_i$ is a fractional $(\kappa, 2^i)$-matching, for some capacity function $\kappa$, and each $x_i$ is contained in $x_{i+1}$.

\begin{definition}[Size factor]
  A fractional $b$-matching $x$ of $G$ has \emph{size factor} $\alpha$ with respect to $\allCj$ if $x$ restricted to $\Ej$ is a $\alpha$-approximate $b$-matching of $\Gj = G[\Cj, S]$ for all $j \in \{0, \dots, \jmax\}$.
\end{definition}

% \begin{remark}
%     When the clients have uniform weights, we only need to compute $(1,
%     B)$-matchings. In this setting, we can apply a slight generalization of the
%     algorithm of Assadi, Liu, and Tarjan~\cite{ALT21} to compute a
%     $(1+\eps)$-approximate $(1, B)$-matching in $O(1/\eps^2)$ passes and
%     $O(n\log{n})$ space.
% \end{remark}

\begin{definition}[Fractional nested matching hierarchy]
    Let $\kappa : C \to \nonneg$.
    Let $x_i : E \to \nonneg$ for $i \in \{0, \dots, \imax\}$.
    A sequence $\bx = (x_0, \dots, x_\imax)$ is a \emph{fractional $(\kappa, 2^i)$-nested matching hierarchy (NMH)} if for all $i \in \{0, \dots, \imax\}$,
    \begin{enumerate}
        \item $x_i$ is a fractional $(\kappa, 2^i)$-matching and
        \item $x_i(e) \ge x_{i-1}(e)$ for all $e \in E$.
    \end{enumerate}
    We say that $\bx$ has \emph{size factor} $\alpha$ with respect to $\allCj$ if each $x_i$ has size factor $\alpha$ with respect to $\allCj$.
\end{definition}

% \begin{remark}
%     The choice of the base of the server capacities in the definition above is likely not optimal for our algorithm, but we have not attempted to optimize the approximation factor.
% \end{remark}

% By running the algorithm from \Cref{cor:sf-alg} in parallel for $i = 0$ to $\imax$, we can compute matchings $x_0, \dots, x_{\imax}$ where $x_i$ is a
% $(\kappa, 2^i)$-matching of size factor $\alpha$.
% But these matchings would not yet form a nested matching hierarchy as the
% matchings would not necessarily contain each other.
% However, by merging the matchings in a greedy fashion similar to \textsc{MergeInto}, we can compute a new set of matchings that are nested and still have large size factor.

Computing a $(w, 2^i)$-NMH of size factor $8$ with respect to $\allCj$ turns out to be relatively simple, even in the semi-streaming setting.
The algorithm boils down to computing $\polylog{n}$ many maximal $b$-matchings.
We defer the details to \Cref{sec:nmh-streaming}.

\subsection{Matching Hierarchies and \texorpdfstring{$\ell_p$}{Lp}-Norms}

Given a fractional $(w, 2^i)$-NMH $\bx = (x_0, \dots, x_{\imax})$, we can obtain a relatively good upper bound on the $\ell_p$-norm of the load vector of the matching $x_{\imax}$ in terms of the other matchings $x_i$.

\begin{lemma}\label{lem:nmh-ub}
  If $\bx = (x_0, \dots, x_{\imax})$ is a fractional $(w, 2^i)$-NMH, then \[
    \sum_{s \in S} x_\imax(\delta(s))^p \le p \cdot x_0(E) + \sum_{i=1}^{\imax} p 2^{i(p-1)} \cdot (x_i - x_{i-1})(E).
  \]
\end{lemma}

\begin{proof}
  Fix $s \in S$.
  Let $d_i = x_i(\delta(s))$ for $i \in \{-1, \dots, \imax\}$.
  Let $g : \R \to \R$ be the function $g(z) = z^p$ and recall that $g'(z) = p z^{p - 1}$.
  Since $g$ is differentiable, \[
    g(x_\imax(\delta(s))) = \sum_{i=0}^{\imax} g(d_i) - g(d_{i-1}) = \sum_{i=0}^\imax \int_{d_{i-1}}^{d_i} g'(z) dz.
  \]
  Using that $g$ is convex and $d_i \le 2^i$, \begin{equation}
    \int_{d_{i-1}}^{d_i} g'(z) dz \le g'(d_i) (d_i - d_{i-1}) \le g'(2^i) (d_i - d_{i-1}).
  \end{equation}
  Hence \[
    x_\imax(\delta(s))^p \le \sum_{i = 0}^\imax p 2^{i(p - 1)} (x_i - x_{i-1})(\delta(s)).
  \]
  Summing over all $s \in S$ gives the result.
\end{proof}

We also prove a complementary lower bound for more structured matching hierarchies.
We first need the following definition.
\begin{definition}
  A fractional $(\kappa, 2^i)$-NMH $(x_0, \dots, x_\imax)$ is \emph{maximally-nested} if $x_i(\delta(s)) = 2^i$ whenever $x_{i+1}(\delta(s)) \ne x_i(\delta(s))$ for all $s \in S$ and $i \in \{0, \dots, \imax - 1\}$.
\end{definition}
Maximally-nested hierarchies are maximal in the sense that for all $i \in \{0, \dots, \imax - 1\}$ and all $e \in E$, one cannot increase $x_i(e)$ without violating the inclusion property of a nested matching hierarchy.
Maximally-nested integral matching hierarchies have the following property, which we utilize in the next proof: for all $i \in \{0, \dots, \imax\}$ and $s \in S$, the quantity $(x_i - x_{i - 1})(\delta(s))$ is either zero, or between $1$ and $2^{i-1}$.

\begin{lemma}\label{lem:nmh-lb}
  If $\bx = (x_0, \dots, x_{\imax})$ is a maximally-nested integral $(w, 2^i)$-NMH, then \[
    \sum_{s \in S} x_\imax(\delta(s))^p \ge x_0(E) + \sum_{i=1}^{\imax} p2^{(i-1)(p-1)} \cdot (x_i - x_{i-1})(E).
  \]
\end{lemma}

\begin{proof}
  The proof is similar to that of \Cref{lem:nmh-ub}, but a bit more delicate.
  Fix $s \in S$ and let $d_i = x_i(\delta(s))$ for $i \in \{0, \dots, \imax\}$.
  Since $g$ is differentiable, \[
    g(x_\imax(\delta(s))) = g(d_0) + \sum_{i=1}^{\imax} g(d_i) - g(d_{i-1})
                          = g(d_0) + \sum_{i=1}^\imax \int_{d_{i-1}}^{d_i} g'(z) dz.
  \]
  Using that $g$ is convex,
  \begin{equation}
    \int_{d_{i-1}}^{d_i} g'(z) dz \ge g'(d_{i-1}) \cdot (d_i - d_{i-1}).
  \end{equation}
  Now since $\bx$ is maximally-nested and $g'$ is increasing, $g'(d_{i-1}) \cdot (d_i - d_{i-1}) \ge g'(2^{i-1}) \cdot (d_i - d_{i-1})$; indeed, either $d_i = d_{i-1}$ or $d_{i-1} = 2^{i-1}$.
  Also, by the integrality of $\bx$, we have $d_0 \in \{0, 1\}$, and in either case $g(d_0) = d_0$.
  We thus have \begin{align*}
    g(x_\imax(\delta(s))) \ge g(d_0) + \sum_{i = 1}^\imax g'(d_{i-1}) \cdot (d_i - d_{i-1})
                          \ge d_0 + \sum_{i = 1}^\imax g'(2^{i-1}) \cdot (d_i - d_{i-1}).
  \end{align*}
  Summing over $s \in S$ yields the desired result.
\end{proof}

We conclude this section with a simple decomposition of an assignment into a maximally-nested matching hierarchy.
This will be useful when we want to apply the lower bound above to an $\ell_p$-norm-minimizing assignment.

\begin{definition}\label{def:decomposition}
  Let $A \subseteq E$ be an assignment and let $\imax = \lceil \log_2(w(C)) \rceil$.
  A $(w, 2^i)$-nested matching hierarchy $\bx = (x_0, \dots, x_\imax)$ is a \emph{$(w, 2^i)$-nested matching hierarchy decomposition} of $A$ if $\bx$ is maximally-nested and $x_{\imax}(cs) = \begin{cases}
      w(c) & cs \in A\\
      0 & \text{otherwise}.
    \end{cases}$
\end{definition}

Every assignment $A$ has a fractional $(w, 2^i)$-NMH decomposition, constructed as follows.
Initialize $x_\imax$ as in the definition and set $x_i(e) = 0$ for all $i < \imax$ and $e \in E$.
Then while there is any $e \in E$ and $i < \imax$ for which $x_i(e) < x_{i+1}(e)$ and $x_i(e) < 2^i$, set $x_i(e) = \min \{ x_{i+1}(e), 2^i \}$.
Once the procedure terminates, the resulting NMH is maximally-nested by definition.

\subsection{An All-Norm Oracle from a Nested Matching Hierarchy}

We now describe how to solve the all-norm oracle problem (\Cref{fig:oracle}) given an algorithm to compute a nested matching hierarchy.
We first compute a fractional $(w, 2^i)$-NMH $\mathbf{y} = (y_0, \dots, y_\imax)$ of size factor $\alpha$ with respect to $\allCj$.
The computation of the NMH $\mathbf{y}$ is the only step that requires access to $G$; the rest of the algorithm is entirely based off of $\mathbf{y}$.

With $\mathbf{y}$ in hand, the algorithm initializes $x(e) = 0$ for all $e \in \supp(y_\imax)$.
The main outer loop of the algorithm increments an iterator $j$ from $0$ to $\jmax$.
In iteration $j$, only those clients in $\Cj$ are considered.
In this way, the outer loop implicitly maintains a large value-per-unit-weight threshold; it starts with the largest threshold producing a non-empty set of clients and decreases it by a factor of $e^{\eps} \approx 1 + \eps$ in each step.

Within each outer loop iteration, an inner loop increments an iterator $i$ over the levels of the matching hierarchy $\mathbf{y} = (y_0, \dots, y_\imax)$.
Within iteration $i$ of the inner loop, the algorithm tries to copy the values of $y_i$ on edges in $\Ej$ into $x$.
However, the algorithm does not allow $x$ to grow too large too quickly; it enforces that $x(\Ej) \le \frac{1}{\alpha} w(\Cj)$ throughout the outer iteration $j$.
More precisely, for any edge $e \in \Ej$ such that $x(e) < y_i(e)$, the algorithm increases $x(e)$ continuously until either $x(e) = y_i(e)$ or $\alpha x(\Ej) = w(\Cj)$.

At the end of the outer loop, computing the final result is just a matter of scaling.
The algorithm sets $z(cs) = (1 + \eps)^2 \alpha x(cs) / w(c)$ for $cs \in \supp(x)$ and returns $z$.
See \Cref{alg:oracle}.

\begin{figure}[ht]
\begin{algorithm}[H]
  \Input{The graph $G = (C, S, E)$, client weights $w : C \to \Z_+$, client values $\lambda : C \to \nonneg$, and an accuracy parameter $\eps > 0$.}
  Set $R \gets \max_{c \in C} \rho(c)$.\;
  Set $\imax \gets \lceil \log_2{n}\rceil$; set $\jmax \gets \lceil \log (R w(C)/\eps)/\eps \rceil$.\;
  \tcp{The quantities $R$ and $\eps$ define the sets $\Cj$.}
  Compute a fractional $(w, 2^i)$-NMH $\mathbf{y} = (y_0, \dots, y_\imax)$ of size factor $\alpha$ w.r.t.\ $\allCj$.\;\label{ln:nmh}
  \tcp{The edges $E$ of the graph are only needed in the previous step to compute $\mathbf{y}$.}
  Set $x(e) \gets 0$ for $e \in \supp(y_\imax)$.\;
  \For{$j \gets 0$ \KwTo $\jmax$}{\label{ln:jloop}
    \For{$i \gets 0$ \KwTo $\imax$}{\label{ln:iloop}
      While $x(\Ej) < \frac{1}{\alpha}w(\Cj)$ and there is some $e \in \Ej$
      such that $x(e) < y_i(e)$, increase $x(e)$ by the smallest amount so that
      either $x(\Ej) = \frac{1}{\alpha}w(\Cj)$ or $x(e) = y_i(e)$.\;
    }
  }
  Set $z(cs) \gets (1 + \eps)^2 \, \alpha \cdot x(cs)/w(c)$ for $cs \in \supp(x)$.\;\label{ln:afterj}
  Return $z$.\;

  \caption{An $O(1)$-all-norm oracle.}\label{alg:oracle}
\end{algorithm}
\end{figure}
\paragraph*{Correctness of the oracle.}
We now begin to prove that \Cref{alg:oracle} solves the all-norm oracle problem (\Cref{fig:oracle}).
There are three things to prove:
\begin{enumerate}
  \item The output $z$ satisfies the combined client-covering constraint \eqref{eq:ccc}.
  \item The output $z$ satisfies the $\beta$-approximate $\ell_p$-norm constraints for some constant $\beta$ \eqref{eq:lpc}.
  \item The output $z$ has width $O(1)$.
\end{enumerate}

We start with a definition regarding \Cref{alg:oracle} and then collect some useful observations, whose proofs are straightforward.

\begin{definition}
  The \emph{outer loop} is the loop beginning on \cref{ln:jloop}, and the \emph{inner loop} is the loop beginning on line \cref{ln:iloop}.
  We say \emph{outer-iteration $j$} to refer to iteration $j$ of the outer loop, and \emph{inner-iteration $(j, i)$} to refer to iteration $i$ of the inner loop during outer-iteration $j$.
\end{definition}

\begin{observation}\label{obs:xjsupport}
  Let $j \in \{0, \dots, \jmax\}$.
  At the end of iteration $j$ of the outer loop, $x$ is supported on $\Ej$.
\end{observation}
%\begin{proof}
%  Each outer-iteration $j$ only increases $x(e)$ for $e \in \Ej$, so by the end of outer-iteration $j$, we have that $x$ is supported on $\Ej[0] \cup \cdots \cup \Ej \subseteq \Ej$.
%\end{proof}

\begin{observation}\label{obs:invariant}
  Let $j \in \{0, \dots, \jmax\}$.
  Throughout iteration $j$ of the outer loop, $x(\Ej) \le \frac{1}{\alpha} w(\Cj).$
\end{observation}
%\begin{proof}
%  The inequality \eqref{eq:invariant} holds trivially at the beginning of
%  outer-iteration $0$.
%  We now observe that whenever \eqref{eq:invariant} holds at the beginning of
%  outer-iteration $j$, it holds throughout outer-iteration $j$.
%  Indeed, any update to $x$ has to pass through \cref{ln:ifcapd,ln:capd}, which
%  ensure that $x(\Ej) \le \frac{1}{\alpha} w(\Cj)$ after the update.
%
%  It remains to show that if $x(\Ej) \le \frac{1}{\alpha} w(\Cj)$ at the end of
%  outer-iteration $j$, then $x(\Ej[j+1]) \le \frac{1}{\alpha} w(\Cj)$ at the
%  beginning of outer-iteration $j+1$.
%  In notation, we assume $\xj[j](\Ej) \le \frac{1}{\alpha} w(\Cj)$ and want to
%  show $\xj(\Ej[j+1]) \le \frac{1}{\alpha} w(\Cj[j+1])$.
%  The impliciation follows from the fact that $\xj$ is supported on $\Ej$ and
%  that $\Cj \subseteq \Cj[j+1]$: \[
%    \xj(\Ej[j+1]) \le \xj(\Ej[j]) \le \frac{1}{\alpha} w(\Cj[j]) \le \frac{1}{\alpha} w(\Cj[j+1]). \qedhere
%  \]
%\end{proof}

\begin{observation}\label{obs:monotonic}
  Throughout the algorithm, $x(e)$ monotonically increases for all $e \in E$.
\end{observation}
%\begin{proof}
%  Let $e \in E$.
%  The only line on which $x(e)$ changes is \cref{ln:setx}, so we just need to
%  argue that the potential values for $d$, assigned on \cref{ln:setd,ln:capd},
%  are nonnegative.
%
%  The if-statement on \cref{ln:if-guard} ensures that $x(e) < y_i(e)$ in every
%  inner-iteration $(j, i, e)$ for which \cref{ln:setx} tries to update $e$, and
%  hence \cref{ln:setd} sets $d$ to $y_i(e) - x(e) > 0$.
%  If $d$ is set again on \cref{ln:capd}, its new value $\frac{1}{\alpha}w(\Cj)
%  - x(\Ej)$ is nonnegative by \Cref{obs:invariant}.
%\end{proof}

\begin{observation}
  At the end of the algorithm, $x(e) \le y_\imax(e)$ for all $e \in E$.
\end{observation}

\begin{observation}\label{obs:saturation}
  At the end of inner-iteration $(j, i)$, either $x(\Ej) = \frac{1}{\alpha} w(\Cj)$ or $x(e) \ge y_i(e)$ for all $e \in \Ej$.
\end{observation}

\begin{observation}
  At all times throughout the algorithm, $x(\delta(c)) \le w(c)$.
\end{observation}

We now show that the inequality in \Cref{obs:invariant} is in fact met with equality at the end of outer-iteration $j$.

\begin{lemma}\label{lem:assertion}
  Let $j \in \{0, \dots, \jmax\}$.
  At the end of iteration $j$ of the outer loop, $x(\Ej) = \frac{1}{\alpha} w(\Cj)$.
\end{lemma}

\begin{proof}
  Let $j \in \{0, \dots, \jmax\}$.
  By \Cref{obs:saturation}, at the end of outer-iteration $j$---that is, at the end of inner-iteration $(j, \imax)$---either $x(\Ej) = \frac{1}{\alpha}w(\Cj)$ or $x(e) \ge y_{\imax}(e)$ for all $e \in \Ej$.
  It therefore suffices to show that $y_{\imax}(\Ej) \ge \frac{1}{\alpha} w(\Cj)$.

  Recall that $y_\imax$ is a $(w, 2^\imax)$-matching of size factor $\alpha$.
  The size of a maximum $(w, 2^\imax)$-matching is easily seen to be $w(C)$.
  Indeed, as $2^\imax = 2^{\lceil \log_2 w(C) \rceil} \ge w(C)$, the server capacities of a $(w, 2^\imax)$-matching are so large that every client can be matched arbitrarily without violating any server capacity.
  Hence $\alpha \cdot y_\imax(\Ej) \ge w(C) \ge w(\Cj).$
\end{proof}

The next lemma establishes that the output of the oracle satisfies the combined client covering constraint.

\begin{lemma}\label{lem:client-constraint}
  The output $z$ of \Cref{alg:oracle} satisfies $\sum_{c \in C} \lambda(c) z(\delta(c)) \ge \sum_{c \in C} \lambda(c).$
\end{lemma}
\begin{proof}
  It suffices to show that at the end of the algorithm
  \begin{equation}\label{eq:mass-prop-conc}
    (1 + \eps) \alpha \sum_{c \in \Cmax} \rho(c) x(\delta(c)) \ge \sum_{c \in \Cmax} \rho(c) w(c).
  \end{equation}
  Indeed, if \eqref{eq:mass-prop-conc} holds, then since $x$ is supported on $\Emax$,
  \begin{align*}
    \sum_{c \in C} \lambda(c) z(\delta(c))
    &= (1 + \eps)^2 \alpha \sum_{c \in C} \rho(c) x(\delta(c))\\
    &= (1 + \eps)^2 \alpha \sum_{c \in \Cmax} \rho(c) x(\delta(c))\\
    &\ge (1 + \eps) \sum_{c \in \Cmax} \rho(c)w(c) = (1 + \eps) \sum_{c \in \Cmax} \lambda(c) \ge \sum_{c \in C} \lambda(c),
  \end{align*}
  where the last inequality is by \Cref{lem:lambda-cmax}.

  So let us prove inequality \eqref{eq:mass-prop-conc}.
  Let $r_j = e^{-\eps j} \cdot \max_{c \in C} \rho(c)$ be the quantity that defines $\Cj$.
  For any function $f : C \to \nonneg$, we can upper- and lower-bound $\sum_{c \in C} f(c) \rho(c)$ by Riemann-like sums:
  \begin{equation}
    \sum_{j = 0}^{\jmax} f(\Cj) (r_{j} - r_{j+1}) \le \sum_{c \in \Cmax} f(c) \rho(c) \le \sum_{j = 0}^{\jmax} f(\Cj) (r_{j-1} - r_{j}).
  \end{equation}
%  and similarly, \[
%    \sum_{c \in \Cmax} \rho(c) x(\delta(c)) \ge \sum_{j = 0}^{\jmax} x(\delta(\Cj)) (r_{j+1} - r_j) = \sum_{j=0}^{\jmax} x(\Ej) (r_{j+1} - r_j).
%  \]
  By \Cref{lem:assertion}, $x(\Ej) = \frac{1}{\alpha} x(\delta(\Cj))$ at the end of iteration $j$ of the outer loop.
  Because $x$ increases monotonically, it follows that at the end of the algorithm, $x(\Ej) \ge \frac{1}{\alpha} x(\delta(\Cj))$ for all $j \in \{0, \dots, \jmax\}$.
  Putting everything together,
  \begin{align*}
    \sum_{c \in \Cmax} w(c) \rho(c) &\le \sum_{j = 0}^{\jmax} w(\Cj) (r_{j-1} - r_{j})\\
                                    &\le \alpha \sum_{j = 0}^{\jmax} x(\Ej) (r_{j-1} - r_{j})\\
                                    &\le (1 + \eps) \alpha \sum_{j = 0}^{\jmax} x(\Ej) (r_{j} - r_{j+1})\\
                                    &\le (1 + \eps) \alpha \sum_{c \in \Cmax} x(\delta(c)) \rho(c). \qedhere
  \end{align*}
\end{proof}

Having shown that the combined client covering constraint is satisfied, we now proceed to showing that the $\ell_p$-norm constraints are satisfied.
One can think of $(x^*_i)_i$ in the lemma below as a $(w, 2^i)$-NMH decomposition of an $\ell_p$-norm optimal assignment (recall \Cref{def:decomposition}), and indeed this is how we apply it later in this subsection.

\begin{lemma}\label{lem:relative-size}
  Let $(x^*_0, \dots, x^*_\imax)$ be a fractional $(w, 2^i)$-NMH decomposition of an assignment and let $x^* = x^*_\imax$.
  At the end of \Cref{alg:oracle}, \[
    \alpha \cdot (x - y_i)^+(E) \le (x^* - x_i^*)(E),
  \]
  for all $i \in \{0, \dots, \imax\}$.
\end{lemma}

\begin{proof}
  \newcommand{\istar}{{i^*}}
  Let $\xj$ denote the value that $x$ takes on at the end of iteration $j$ of the outer loop.
  Equivalently, $\xj$ is the value that $x$ takes on at the beginning of iteration $j+1$ of the outer loop, and so we also define $\xj[-1]$ to be the zero vector.

  Let $P(j)$ be the following proposition:
  \begin{equation}
    \alpha \cdot (\xj - y_i)^+(\Ej) \le (x^* - x_i^*)(\Ej) \text{ for all $i \in \{0, \dots, \imax\}$.}
  \end{equation}

  We show that $P(j)$ holds for all $j \in \{-1, \dots, \jmax\}$ by induction.
  But first, let us see why $P(\jmax)$ implies the statement of the lemma.
  The value of $x$ at the end of the algorithm is also known as $\xmax$, and so since $P(\jmax)$ holds, we have \[
    \alpha \cdot (\xmax - y_i)^+(\Emax) \le (x^* - x_i^*)(\Emax)
  \]
  for all $i$.
  The quantity on the left is the same as $\alpha \cdot (\xmax - y_i)^+(E)$ because $\xmax$ is supported on $\Emax$.
  The quantity on the right is trivially at most $(x^* - x_i^*)(E)$ since $\Emax \subseteq E$.

  Now we return to proving $P(j)$ for all $j \in \{-1, \dots, \jmax\}$.
  Fix $j$.
  If $j = -1$, then the claim holds trivially.
  Otherwise, $j \ge 0$ and we may assume inductively that $P(j - 1)$ holds.
  During outer-iteration $j$, there are two types of inner-iterations $i$: those that end with $x(\Ej) < \frac{1}{\alpha}w(\Cj)$ (and hence $x \ge y_i$) and those that end because $x(\Ej) = \frac{1}{\alpha} w(\Ej)$.

  \textbf{Case 1}: $x(\Ej) < \frac{1}{\alpha} w(\Cj)$ at the end of inner-iteration $(j, i)$.
  We do not need the inductive hypothesis for this case.
  At the end of inner-iteration $(j, i)$, since $x(\Ej) < \frac{1}{\alpha} w(\Cj)$, it must be that $x(e) \ge y_i(e)$ for all $e \in \Ej$ (or else the iteration would not be complete).
  By monotonicity, $\xj(e) \ge y_i(e)$ for all $e \in \Ej$.
  Hence \[
    \alpha \cdot (\xj - y_i)^+(\Ej) = \alpha \cdot (\xj - y_i)(\Ej) = \alpha \xj(\Ej) - \alpha y_i(\Ej).
  \]
  By \Cref{lem:assertion}, we have $\alpha \xj(\Ej) = w(\Cj)$, and $w(\Cj) = x^*(\Ej)$ by the definition of $x^*$ (see \Cref{def:decomposition}).
  Lastly, observe that $\alpha y_i(\Ej) \ge x^*_i(\Ej)$ because $y_i$ is a $(w, 2^i)$-matching of size factor $\alpha$ and $x_i^*$ is a $(w, 2^i)$-matching.

  \textbf{Case 2}: $x(\Ej) = \frac{1}{\alpha} w(\Cj)$ at the end of inner-iteration $(j, i)$.
  We argue that
  \begin{equation}\label{eq:unchanged}
    (\xj - y_i)^+(e) = (\xj[j-1] - y_i)^+(e) \text{ for all $e \in \Ej$.}
  \end{equation}
  Let us first see why this is sufficient to establish $P(j)$, assuming $P(j - 1)$.
  We have
  \begin{align*}
    (\xj - y_i)^+(\Ej) &= (\xj[j-1] - y_i)^+(\Ej) && (\text{by \eqref{eq:unchanged}})\\
                       &= (\xj[j-1] - y_i)^+(\Ej[j-1]) && (\text{because $\xj[j-1]$ is supported on $\Ej[j-1]$})\\
                       &\le (x^* - x_i^*)(\Ej[j-1]) && (\text{by $P(j-1)$})\\
                       &\le (x^* - x_i^*)(\Ej).
  \end{align*}

  Now let us prove \eqref{eq:unchanged}.
  Let $e \in \Ej$.
  If $x(e)$ does not increase at all in outer-iteration $j$, then we are done.
  So assume $x(e)$ increases in outer-iteration $j$ and let $\istar$ be the largest integer such that $x(e)$ increases in inner-iteration $(j, \istar)$.
  Since at the end of inner-iteration $(j, i)$ we have $x(\Ej) = \frac{1}{\alpha} w(\Cj)$ (by assumption), $x$ does not increase for any inner-iteration $(j, i')$ with $i' > i$.
  We conclude that $\istar \le i$.

  Since inner-iteration $(j, \istar)$ is the last inner-iteration of outer-iteration $j$ to increase $x(e)$, and it increases $x(e)$ to at most $y_\istar(e)$, we have $\xj(e) \le y_\istar(e)$.
  Since $\mathbf{y}$ is a nested matching hierarchy and $\istar \le i$, we have $y_\istar(e) \le y_i(e)$.
  Hence $(\xj - y_i)^+(e) = 0$, and by monotonicity, $(\xj[j-1] - y_i)^+(e) = 0$ too.
\end{proof}

%  It remains to prove that \eqref{eq:unchanged} holds for all $e \in \Ej$.
%  If $\xj(e) \le y_i(e)$, then by monotonicity $\xj[j-1](e) \le y_i(e)$, and
%  hence both $(\xj - y_i)^+(e)$ and $(\xj[j-1] - y_i)^+(e)$ are zero.
%
%  If on the other hand $\xj(e) > y_i(e)$, then we claim that $\xj(e) =
%  \xj[j-1](e)$, which implies \eqref{eq:unchanged} trivially.
%  Indeed, by the definition of $\istar$, at the end of inner iteration $\istar$
%  we have $x(\Ej) = \frac{1}{\alpha}(\Ej)$.
%  This means that all subsequent inner iterations (in this iteration of the
%  outer loop) have no effect on $x$ since the termination condition will be satisfied
%  at the beginning of the iteration.
%
%  Therefore only inner iterations $i'$ that can increase $x(e)$ are in $\{0,
%  \dots, \istar\}$. and because $x(e)$ is increased to at most $y_{i'}
%
%  Therefore the largest that $x(e)$ can be \emph{increased} to during outer
%  iteration $j$ is $y_\istar(e)$.
%  Since $i \ge \istar$ and $\mathbf{y}$ is a nested matching hierarchy, $y_i(e) \ge y_\istar(e)$.
%  Hence if $x(e) > y_i(e) \ge y_\istar(e)$ at the end of outer iteration $j$, it
%  must be that $x(e)$ did not increase during outer iteration $j$, and so
%  $\xj(e) = \xj[j-1](e)$.

\begin{lemma}\label{lem:lp-norm-constraints}
  Let $z$ be the output of \Cref{alg:oracle}. For all $p \ge 1$, \[
    \lpcostz \le \frac{2^{1-1/p}}{\alpha^{1/p}} \cdot \OPT_p.
  \]
\end{lemma}

\begin{proof}
  Fix $p \ge 1$.
  At the end of the algorithm, we have \[
    \sum_{s \in S} \left( \sum_{c \in N(s)} w(c) z(\delta(s))\right)^p = \sum_{s \in S} x(\delta(s))^p.
  \]

  Define $x_i = \min \{ x, y_i \}$ for $i \in \{0, \dots, \imax\}$.
  To make the notation more manageable, let us also define $a_i = p 2^{i(p - 1)}$ for $i \in \{0, \dots, \imax\}$.
  By \Cref{lem:nmh-ub},
  \begin{align*}
    \sum_{s \in S} x(\delta(s))^p &\le a_0 x_0(E) + \sum_{i=1}^\imax a_i (x_i - x_{i - 1})(E)\\
                                  &= a_0 x_\imax(E) + \sum_{i=1}^\imax (a_i - a_{i-1}) (x_\imax - x_{i-1})(E).
  \end{align*}
  Now let $A^*$ be the $\ell_p$-norm-minimizing assignment and let $\bx^* = (x_0^*, \dots, x^*_\imax)$ be a max\-imally-nested $(w, 2^i)$-NMH decomposition for $A^*$.
  By \Cref{lem:nmh-lb},
  \begin{align*}
    \sum_{s \in S} x^*(\delta(s))^p &\ge x^*_0(E) + \sum_{i=1}^\imax a_{i-1} (x^*_i - x^*_{i - 1})(E)\\
                                    &= x^*_\imax(E) + (a_0 - 1)(x^*_\imax - x_0)(E) + \sum_{i=2}^\imax (a_{i-1} - a_{i-2}) (x^*_\imax - x^*_{i - 1})(E).
  \end{align*}

  The upper and lower bounds nearly match; we can use \Cref{lem:relative-size} to transform one into the other.
  Recall that \Cref{lem:relative-size} states that \[
    \alpha \cdot (x_\imax - x_{i - 1})(E) \le (x^*_\imax - x^*_{i - 1})(E).
  \]
  Hence \begin{align*}
    \sum_{s \in S} x(\delta(s))^p &\le a_0 x_\imax(E) + \sum_{i=1}^\imax (a_i - a_{i-1}) \cdot (x_\imax - x_{i-1})(E)\\
                                  &\le \frac{1}{\alpha} \left(a_0 x^*_\imax(E) + \sum_{i=1}^\imax (a_i - a_{i-1}) (x^*_\imax - x^*_{i-1})(E)\right)\\
                                  &\le \frac{2^p}{\alpha} \left(x^*_\imax(E) + (a_0 - 1)(x^*_\imax - x_0)(E) + \sum_{i=2}^\imax (a_{i-1} - a_{i-2}) (x^*_\imax - x^*_{i-1})(E)\right)\\
                                  &\le \frac{2^p}{\alpha} \sum_{s \in S} x^*(\delta(s))^p = \frac{2^p}{\alpha} \cdot \OPT_p^p.\qedhere
  \end{align*}
\end{proof}

\begin{lemma}\label{lem:oracle}
  For every $\eps > 0$, \Cref{alg:oracle} produces a solution to the $(1 + \eps)^2 2\alpha^{1 - 1/p}$-all-norm oracle problem.
\end{lemma}
\begin{proof}
  Let $z$ be the output of \Cref{alg:oracle} and let $x$ be the quantity computed by the oracle at the end of the algorithm.
  Since $z(cs) = (1 + \eps)^2 \alpha x(cs) / w(c)$ for $cs \in \supp(x)$, we have \[
    \sum_{c \in C} \lambda(c) z(\delta(c)) = (1 + \eps)^2 \alpha \sum_{c \in C} \rho(c) x(\delta(c)).
  \]

  \Cref{lem:client-constraint} then says precisely that the combined-client covering constraint is satisfied. By \Cref{lem:lp-norm-constraints},
  \begin{align*}
    \lpcostz &= (1 + \eps)^2 \alpha \cdot \left(\sum_{s \in S} x(\delta(s))^p \right)^{1/p}\\
      &\le (1 + \eps)^2 (2 \alpha)^{1 - 1/p} \cdot \OPT_p.
  \end{align*}
  Finally, since $x(\delta(c)) \le w(c)$ for all $c \in C$, it follows that $z(\delta(c)) \le (1 + \eps)^2 \alpha$ for all $c \in C$.
\end{proof}

\begin{lemma}\label{lem:oracle-streaming}
  If there is a semi-streaming algorithm to compute a $(w, 2^i)$-NMH of size factor $\alpha$ using $p$ passes, then there is a semi-streaming algorithm to compute a $(1 + \eps)2\alpha$-all-norm fractional assignment.
\end{lemma}

\begin{proof}
  Run \Cref{alg:oracle} using the semi-streaming algorithm to implement \cref{ln:nmh}.
  The NMH $\mathbf{y}$ output must have a support of size $O(n \polylog{n})$.
  The additional space used by \Cref{alg:oracle} is in computing $x$ and $z$, which have the same support as $y_\ell$.
  Hence the entire algorithm uses $O(n \polylog{n})$ space.

  By \Cref{lem:oracle}, the output is a $(1+\eps)(2\alpha + 1)$-all-norm assignment.
\end{proof}

\subsection{Nested Matching Hierarchies in the Semi-Streaming Model}\label{sec:nmh-streaming}

We have reduced our task to computing a nested matching hierarchy with respect to $\allCj$ in the semi-streaming setting.

\paragraph*{Computing $b$-matchings of size factor $4$.}
Crouch and Stubbs~\cite{CrouchStubbs14} described a single-pass algorithm to compute a simple matching (i.e., integral and unit-capacitated) of size factor $2\alpha$ using any $\alpha$-approximate maximum cardinality matching algorithm.
Their algorithm generalizes to the fractional $b$-matching setting, and hence any $\alpha$-approximate maximum cardinality $b$-matching algorithm can be used to compute a $b$-matching with size factor $2\alpha$.

\SetKwFunction{MergeInto}{MergeInto}

The main subroutine is a greedy algorithm $\MergeInto{x, y}$, which takes as input two $b$-matchings $x$ and $y$ and returns a $b$-matching $\tx$ that (1) contains $x$ and (2) has size at least half that of the size of $y$.
The algorithm works by simply adding edges greedily (and possibly fractionally) from $y$ into $x$ subject to the capacity constraints $b$.

\begin{lemma}\label{lem:merge-size}
  Let $x$ and $y$ be $b$-matchings.
  The output $\tx \gets \MergeInto{x, y}$ is a $b$-matching satisfying $\tx(e) \ge x(e)$ for all $e \in E$ and $2\tx(E) \ge y(E)$.
\end{lemma}
We omit the proof of the previous lemma, which is similar to the proof that the size of a maximal matching is at least half the size of a maximum matching.

We use the greedy \MergeInto procedure in conjunction with the $\alpha$-approximate maximum cardinality matching algorithm to compute a $b$-matching with size factor $2\alpha$.
The algorithm works as follows.
First it computes an $\alpha$ approximate $b$-matching $\xj$ for each subgraph $\Gj$.
It then sets $\txj[1] \gets \xj[1]$, and for each $j \ge 1$, it sets $\txj \gets \MergeInto{\txj[j-1], \xj}.$
Lastly, it returns $\txmax$.
The algorithm is summarized in \Cref{alg:size-factor-matching}.

\begin{figure}[ht]
\begin{algorithm}[H]
  \Input{An iterator over the edges of $G$, client capacities $\kappa$, and server capacities $\tau$.}
  \Output{A $(\kappa, \tau)$-matching of size factor 4.}

  \proc{\MergeInto{x, y}}{
    Set $\tx(e) \gets x(e)$ for each $e \in \supp(x) \cup \supp(y)$.\;
    \For{$cs \in \supp(x)$}{
      Increase $\tx(cs)$ by the smallest amount such that either
      $\tx(\delta(c)) = \kappa(c)$, $\tx(\delta(s)) = \tau(s)$, or $\tx(cs) =
      x(cs) + y(cs)$.\;
    }
    Return $\tx$.\;
  }

  Compute a maximal $(\kappa, \tau)$-matching $\xj$ of $\Gj$ for each $j \in \{0, \dots,
  \jmax\}$ using one pass over the stream of edges.\label{ln:maximal}\;
  Set $\txj[0] \gets \xj[0]$.\;
  \lFor{$j \gets 1$ \KwTo $\jmax$}{
    set $\txj \gets \MergeInto(\txj[j-1], \xj)$.\label{ln:merge}
  }
  Return $\txmax$.\;

  \caption{Computing a $(\kappa, \tau)$-matching of size factor 4.}\label{alg:size-factor-matching}
\end{algorithm}
\end{figure}

\begin{lemma}
  The value $\txmax$ returned by \Cref{alg:size-factor-matching} is a
  $b$-matching of size factor $2\alpha$.
\end{lemma}
\begin{proof}
  Let $j \in \{0, \dots, \jmax\}$ and let $y$ be a $b$-matching.
  By \Cref{lem:merge-size} and a simple induction, $\txmax$ contains $\txj$.
  Now we have
  \begin{align*}
	  2\alpha \txj(\Ej) &\ge \alpha \xj(\Ej) && (\text{by \Cref{lem:merge-size}})\\
						&= \alpha \xj(E) && (\supp(\xj) \subseteq \Ej)\\
						&\ge y(\Ej). && (\text{$\xj$ is $\alpha$-approximate in $\Gj$}) \qedhere
  \end{align*}
\end{proof}

\begin{lemma}\label{lem:size-factor-matching}
    If there is a $p$-pass $S(n, m)$-space streaming algorithm to compute an $\alpha$-approximate $b$-matching, then there is a $p$-pass $O(S(n, m) \cdot \jmax)$-space streaming algorithm to compute a $b$-matching with size factor $2\alpha$ with respect to $\allCj$.
\end{lemma}

\begin{proof}
  The first step of \Cref{alg:size-factor-matching}, \cref{ln:maximal}, can be implemented in the semi-streaming setting using $p$ passes as follows.
  The algorithm first instantiates a streaming algorithm $A_j$ corresponding to $\Gj$ for each $j \in \{0, \dots, \jmax\}$.
  On the arrival of each edge $e$ (over the $p$ passes of the stream), it sends $e$ to each $A_j$ for which $e \in \Gj$.
  At the end of the $p$ passes, each algorithm $A_j$ returns $\xj$.

  The next step of the algorithm, the merge step, is computed directly from $\xj$ and requires no additional passes over the stream.
  The total space used is $O(S(n, m) \cdot \jmax)$.
\end{proof}

Just like a maximal matching can be computed in a single pass and is 2-approximate, a maximal fractional $b$-matching can be computed in a single pass and is 2-approximate.
\begin{lemma}\label{lem:greedy-matching}
  There is a single-pass $O(n\log{n})$-space algorithm to compute a $2$-approximate fractional $b$-matching.
\end{lemma}
\begin{proof}
  The algorithm maintains a fractional $b$-matching $x$, initially empty.
  On the arrival of each edge $e$, it sets $x(e)$ to be as large as possible subject to the capacity constraints.
  If $x(e)$ is changed at all, this step always saturates one of the endpoints.
  Hence at the end of the algorithm, the support of $x$ is $O(n)$.

  The fact that $x$ is a 2-approximation is folklore, and also follows from \Cref{lem:merge-size}: If $x^*$ is a maximum fractional $b$-matching, the output $\tx$ of $\textsc{MergeSize}(x, x^*)$ contains $x$ and has size at least half of $x^*$.
  Since $x$ is maximal, $\tx$ must be equal to $x$.
\end{proof}

As a simple corollary of \Cref{lem:size-factor-matching,lem:greedy-matching}, we can compute a $b$-matching of size factor $4$ in one pass.
As mentioned earlier, this analogous to the algorithm of Crouch and Stubbs~\cite{CrouchStubbs14}, but for fractional $b$-matchings rather than simple matchings.

\begin{corollary}\label{cor:sf-alg}
    There is a single-pass $O(n\log{n} \cdot \jmax)$-space streaming algorithm to compute a $b$-matching with size factor $4$.
\end{corollary}

\SetKwFunction{HMergeInto}{HMergeInto}

We now explain how to compute a fractional $(\kappa, 2^i)$-NMH of size factor $2\alpha$ using an algorithm to compute $b$-matchings of size factor $\alpha$.
The basic subroutine, \HMergeInto, takes as input a $(\kappa, \tau)$-matching $x$ and a $(\kappa, 2\tau)$-matching $y$ and returns a $(\kappa, 2\tau)$-matching $\tx$ such that (1) $\tx$ contains $x$ and (2) $2\tx(\Ej) \ge y(\Ej)$ for all $j \in \{0, \dots, \jmax\}$.
The algorithm works much like \textsc{MergeInto}, except it tries to merge the edges of $y$ in decreasing order of density; see \Cref{alg:nmh}.
The next lemma summarizes the key properties of \HMergeInto.

\begin{lemma}\label{lem:hmerge}
    Let $x$ be a fractional $(\kappa, \tau)$-matching let $y$ be a fractional $(\kappa, 2 \tau)$-matching.
    The output $\tx$ of \HMergeInto{x, y, $2\tau$} satisfies
    \begin{enumerate}
        \item $\tx$ is a $(\kappa, 2\tau)$-matching
        \item $\tx$ contains $x$
        \item $2\tx(\Ej) \ge y(\Ej)$ for all $j \in \{0, \dots, \jmax\}$.
    \end{enumerate}
    In particular, if $y$ has size factor $\alpha$ then $\tx$ has size factor $2\alpha$.
\end{lemma}

The proof \Cref{lem:hmerge} is routine, though somewhat tedious. We defer it to the full version of the paper.

\begin{figure}[ht]
\begin{algorithm}[H]
  \Input{An iterator over the edges of $G$, client capacities $\kappa$, and server capacities $\tau$.}
  \Output{A $(\kappa, 2^i)$-nested-matching hierarchy of size factor 4.}

  \proc{\HMergeInto{x, y, $\tau$}}{
    Set $\tx(e) \gets x(e)$ for each $e \in \supp(x) \cup \supp(y)$.\;
    \For{$j \gets 0$ \KwTo $\jmax$}{
      \For{$cs \in \supp(y) \cap (\Ej \setminus \Ej[j-1])$}{
        Increase $\tx(cs)$ by the smallest amount such that either
        $\tx(\delta(c)) = \kappa(c)$, $\tx(\delta(s)) = \tau(s)$, or $\tx(cs) =
        x(cs) + y(cs)$.\;
      }
    }
    Return $\tx$.\;
  }
  Compute a $(\kappa, 2^i)$-matching $x_i$ with size factor $4$ for $i \in \{0, \dots, \imax\}$ using one pass over the stream of edges.\label{ln:size-factor}\;
  Set $\tx_0 \gets x_0$.\;
  \lFor{$i \gets 1$ \KwTo $\imax$}{
    set $\tx_i \gets \HMergeInto(\tx_{i-1}, x_i, 2^i)$.
  }
  Return $\bx = (\tx_0, \dots, \tx_\imax)$.\;

  \caption{Computing a $(\kappa, 2^i)$-NMH of size factor 8.}\label{alg:nmh}
\end{algorithm}
\end{figure}

The next lemma proves that \Cref{alg:nmh} is correct and can be implemented in in the semi-streaming setting.

\begin{lemma}
    Let $\kappa : C \to \mathbb{R}_+$.
    If there is a $p$-pass $S(n, m)$-space streaming algorithm to compute a fractional $b$-matching of size factor $\alpha$, then there is a $p$-pass $O(S(n, m)\cdot \imax)$-space streaming algorithm to compute a fractional $(\kappa, 2^i)$-NMH of size factor $2\alpha$ with respect to $\allCj$.
\end{lemma}

\begin{proof}
    Using the given algorithm, takes $p$ passes and $O(S(n, m) \cdot \imax)$ space to run \cref{ln:size-factor} of \Cref{alg:nmh}.
    The other steps do not require passes over the stream and the additional space used is also $O(S(n, m) \cdot \imax)$.

    It remains to show that the value $\mathbf{\tx}$ returned by the stream is a fractional $(\kappa, 2^i)$-NMH of size factor $2\alpha$, which follows quickly from \Cref{lem:hmerge}.
    First we have that $\tx_0$ is equal to $x_0$ and hence $\tx_0$ is a fractional $(1, 1)$-matching of size factor $\alpha$.
    For $i \in \{1, \dots, \imax \}$, we have $\tx_i = \HMergeInto(\tx_{i-1}, x_i, 2^i)$, and so \Cref{lem:hmerge} states that $\tx_i$ contains $\tx_{i-1}$ and $\tx_i$ has size factor $2\alpha$.
    Hence $(\tx_0, \dots, \tx_\imax)$ is a fractional $(1, 2^i)$-NMH.
\end{proof}

The upshot from this subsection is the following corollary.

\begin{corollary}\label{lem:compute-nmh}
    There is a single-pass $O(n \cdot \imax \cdot \jmax)$-space streaming algorithm to compute a fractional $(\kappa, 2^i)$-NMH $(x_0, \dots, x_{\imax})$ of size factor $8$ with respect to $\allCj$.
\end{corollary}

We conclude with our main theorem.
\begin{theorem}\label{thm:main}
  There exists a $O(\log{n})$-pass semi-streaming algorithm to compute an $19$-all-norm assignment.
\end{theorem}
\begin{proof}
  By \Cref{lem:compute-nmh}, there is a single-pass semi-streaming algorithm to compute a fractional $(w, 2^i)$-NMH of size factor 8 with respect to $\allCj$.
  By choosing $\eps = 1/16$, it follows by \Cref{lem:oracle-streaming} that there is a semi-streaming algorithm to compute a $17$-all-norm fractional assignment.
  This in turn implies by \Cref{lem:oracle-reduction} that there is $O(\log{n})$-pass a semi-streaming algorithm to compute a $19$-all-norm assignment.
\end{proof}

\bibliography{main}

\end{document}